# Robust Diamond Quantum Sensing in Double-Quantum 4-Ramsey Interferometry by Dual-Feedback Scheme

Yuta Araki,[1] Takeharu Sekiguchi,[1] Yuji Hatano,[1] Naota Sekiguchi,[1] Zehan Li,[1] Chikara Shinei,[2,3] Masashi Miyakawa,[4] Takashi Taniguchi,[4] Tokuyuki Teraji,[2] Hiroshi Abe,[5] Shinobu Onoda,[5] Takeshi Ohshima,[5,6] Mutsuko Hatano,[1] and Takayuki Iwasaki[1]

[1] *Department of Electrical and Electronic Engineering, Institute of Science Tokyo, Meguro, Tokyo 152-8550, Japan*

[2] *Research Center for Electronic and Optical Materials, National Institute for Materials Science, Tsukuba, Ibaraki 305-0044, Japan*

[3] *Department of Applied Physics, University of Tsukuba, Tsukuba, Ibaraki 305-8571, Japan*

[4] *Research Center for Materials Nanoarchitectonics, National Institute for Materials Science, Tsukuba, Ibaraki 305-0044, Japan*

[5] *Takasaki Institute for Advanced Quantum Science, National Institutes for Quantum Science and Technology : QST, Takasaki, Gunma 370-1292, Japan*

[6] *Department of Materials Science, Tohoku University, Sendai, Miyagi 980-8579, Japan*

We developed a highly sensitive and long-term stable diamond quantum magnetometer by integrating a dual-feedback system that compensates for common- and differential-mode frequency shifts in double-quantum (DQ) 4-Ramsey interferometry. Detecting extremely weak magnetic-field signals requires extensive signal accumulation to enhance the signal-to-noise ratio. However, deviations from the optimal operating frequencies caused by thermal drift or environmental magnetic noise severely hinder the long-term stability of such a sensing system. To overcome the limitation, we first derived analytical solutions for the signal response to common- and differential-mode detunings based on a detuning-inclusive Hamiltonian to theoretically clarify the robustness limit in DQ 4-Ramsey sensing. Guided by this theoretical model, we introduced synchronized frequency modulation and a dual-feedback scheme to independently track and compensate for these frequency shifts while maintaining an optimal magnetic-field response. The proposed method enabled stable continuous operation against temperature variations of 2.8 K. By maintaining a magnetic field sensitivity of 7.0 $\mathrm{pT}/\sqrt{\mathrm{Hz}}$ in the frequency range of $5-100$ Hz for a prolonged duration of 4 hours, we achieved an extremely low noise floor of 64 $\mathrm{fT}/\sqrt{\mathrm{Hz}}$. During prolonged continuous sensing, the system exhibited the expected $1/\sqrt{T}$ scaling with integration time $T$, and prevented both the underestimation of amplitude and waveform distortion. These results establish a robust approach for long-term DQ 4-Ramsey diamond magnetometry and demonstrate its potential for detecting weak biomagnetic signals.

## I. Introduction

Nitrogen-vacancy (NV) centers in diamond have emerged as highly promising quantum sensors for biomagnetic field measurements, offering high sensitivity and spatial resolution [1]. Recently, biomagnetic field measurements have been demonstrated, such as human magnetocardiography (MCG) [2], rat cardiac signals [3,4], and other localized biomagnetic signals from ex vivo tissues [5,6] with ensemble NV centers using continuous-wave optically detected magnetic resonance (CW-ODMR). However, detecting even weaker sub-picotesla signals, such as those generated in evoked human magnetoencephalography (MEG) ($\approx 100$ fT [7,8]), necessitates further improvement in intrinsic magnetic field sensitivity.

Ramsey-based magnetometry offers higher magnetic field sensitivity than CW-ODMR due to longer spin dephasing time, avoiding optical and MW power broadening [9]. To further enhance the sensitivity of Ramsey-based sensing, protocols exploiting the double-quantum (DQ) basis ($|m_s=+1\rangle$ and $|m_s=-1\rangle$) have been developed [10]. Specifically, the DQ 4-Ramsey sequence [11] cancels out the dependence of spin state evolution on the zero-field splitting, thereby providing immunity against phase errors induced by temperature fluctuations ($-74$ kHz/K [12]) and strain inhomogeneity. Having practically demonstrated sensitivities superior to CW-ODMR, it is increasingly being utilized for highly sensitive near-DC magnetic field measurements [13,14].

However, for practical biomagnetic sensing using the DQ 4-Ramsey sequence, it is challenging to maintain the high-sensitivity against external disturbances over extended periods. The minimum detectable magnetic field scales as $\sigma_B \approx \eta/\sqrt{T}$, where $\sigma_B$, $\eta$, and $T$ denote the magnetic uncertainty, magnetic field sensitivity, and total measurement duration, respectively [15,16]. During prolonged measurements, thermal drifts (e.g., microwave (MW) and radiowave (RF) heat generation [17]) and environmental magnetic-field drifts (e.g., the Earth's magnetic field) inevitably occur. Ideally, the DQ 4-Ramsey sequence perfectly cancels out temperature fluctuation, assuming perfect superposition states are prepared under zero detuning. In reality, state preparation uses MW pulses of finite duration. Any detuning present during these finite pulses degrades the superposition state and drives the system away from the exact resonance point, resulting in the lowering of the sensitivity and distortion of the measured signal.

In CW-ODMR, proportional-integral-derivative (PID) control is widely used to actively track the resonance frequency [18,19,20,21]. However, similar PID techniques cannot be directly applied to pulsed measurements. Stabilization of pulsed measurements has also been demonstrated using nuclear-spin transitions [22,23], but such approaches are not readily transferable to electronic-spin-based DQ 4-Ramsey magnetometry. In particular, the detailed response of the DQ 4-Ramsey magnetometry to detuning has not been comprehensively formulated, and no systematic compensation method has been established. Thus, simultaneously achieving high magnetic field sensitivity and long-term stability within the DQ 4-Ramsey framework remains challenging.

In this study, we theoretically model and experimentally characterize the response of DQ 4-Ramsey magnetometry to common- and differential-mode detunings and establish a framework for independent tracking. We realize highly sensitive and robust magnetic-field sensing, compensating for the resonance-frequency drifts by introducing synchronized frequency modulation and a dual-feedback scheme compatible with the DQ 4-Ramsey sequence. The proposed method maintains the optimal magnetic-field response under temperature variations up to 2.8 K, while actively compensating the associated resonance-frequency shifts.

Furthermore, by maintaining a high magnetic field sensitivity of $7.0\ \mathrm{pT}/\sqrt{\mathrm{Hz}}$ in the frequency range of $5-100$ Hz for an extended duration of 4 hours using the dual-feedback scheme, we achieved a low noise floor of $64\ \mathrm{fT}/\sqrt{\mathrm{Hz}}$. During prolonged continuous sensing over 4 hours, the magnetic root-mean-square (RMS) noise exhibited the expected $1/\sqrt{T}$ scaling, and the system effectively prevented both the underestimation of signal amplitude and waveform distortion. These results establish a robust approach for long-term diamond quantum magnetometry and highlight its potential for highly sensitive biomagnetic sensing.

## II. Theory of Dynamic Frequency Tracking in DQ 4-Ramsey Magnetometry

### A. Theoretical Formulation

In this section, we derive analytical solutions for the signal response of the DQ 4-Ramsey sequence to common- and differential-mode detunings, accounting for MW pulse imperfections arising from detuning over finite pulse durations. The Hamiltonian of the $NV^-$ ground state is expressed in angular frequency units as [24]

$$H_0 \approx 2\pi D S_z^2 + \gamma_e B_z S_z, \qquad (1)$$

where $S_z$ is the dimensionless spin-1 operator along the NV axis, $D$ is the temperature-dependent zero-field splitting (ZFS), $B_z$ is the projection of the external magnetic field along the NV axis (as shown in Fig. 1(a)), and $\gamma_e = 2\pi \times 28.03$ GHz/T is the $NV^-$ gyromagnetic ratio.

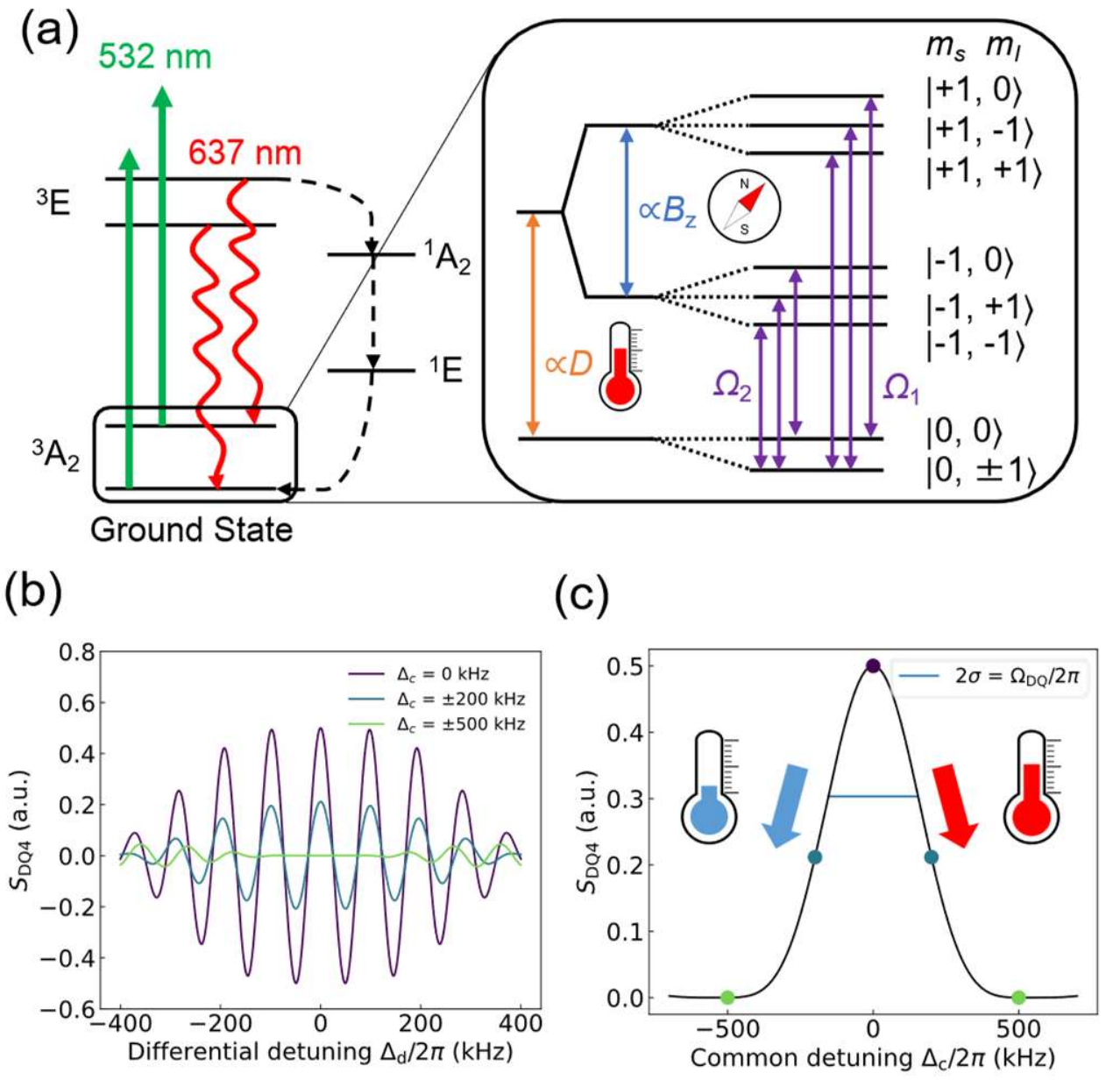

FIG 1. (a) The energy level diagram of $NV^-$ ground state, illustrating the splitting due to magnetic field and temperature. Analytical solutions of the DQ 4-Ramsey signal for (b) differential-mode detuning $\Delta_d$ and (c) common-mode detuning $\Delta_c$. $\Omega_{DQ}/2\pi$ is set at 310 kHz. The amplitude decreases for temperature shifts in both ways and almost vanishes for $|\Delta_c|/2\pi > 400$ kHz ($\pm 5$ K in temperature).

Here, we introduce the two-tone MW driving Hamiltonian [24]:

$$H_{1+2} = \sum_{j=1}^{2} \gamma_e B_j S_x \cos(\omega_j t + \phi_j), (2)$$

where, $B_j$, $\omega_j$, and $\phi_j$ denote the magnetic field amplitude, angular frequency, and phase of the $j$-th MW tone ($j$=1,2) driving the $|0\rangle \leftrightarrow |+1\rangle$ and $|0\rangle \leftrightarrow |-1\rangle$ transitions, respectively. By transforming into the rotating frame and applying the rotating-wave approximation (RWA), the spin evolution under the DQ 4-Ramsey protocol can be analytically formulated. Let $\Delta_1$ and $\Delta_2$ be the angular frequency detunings of the two MW tones from their respective resonances. The differential-mode detuning $\Delta_d$ and common-mode detuning $\Delta_c$ are defined as

$$\Delta_d \equiv \frac{\Delta_1 - \Delta_2}{2}, \qquad \Delta_c \equiv \frac{\Delta_1 + \Delta_2}{2}. \ (2')$$

The corresponding phases are similarly defined as $\phi_\mathrm{d} = (\phi_1 - \phi_2)/2$ and $\phi_\mathrm{c} = (\phi_1 + \phi_2)/2$. After a free-precession time $\tau$, the signal response of a single DQ Ramsey block, $S_{\mathrm{DQ1}}^{(\phi_1,\phi_2)}$, with alternating second-pulse phases $(\phi_1, \phi_2)$, is expressed as the sum of distinct interferometric components: DC, DQ, and residual SQ;

$$S_{\mathrm{DQ1}}^{(\phi_1,\phi_2)} = c_0 + a_{2\mathrm{d}} \cos(2\Delta_\mathrm{d}\tau + 2\phi_\mathrm{d} - 2\psi_\mathrm{d}) + \sum_{j=1}^{2} a_j \cos(\Delta_j \tau + \phi_j - \psi_j), (3)$$

where $\psi_\mathrm{d}$ and $\psi_j$ are the parasitic phase shifts owing to pulse imperfections during the finite MW duration $t_\mathrm{p}$. The detailed derivations and the expressions of involved parameters are provided in Appendix A.

Table I summarizes the signs of the DC, DQ, and SQ signal components under conventional MW phase combinations labeled A, B, C, and D (Fig. 2(a)). Table II details the corresponding phase-cycling combinations. Specifically, the phase-cycling combination of $(\mathrm{A} - \mathrm{B} + \mathrm{C} - \mathrm{D})/4$ cancels out the DC baseline and SQ leakage, yielding the standard DQ 4-Ramsey signal [11]:

$$S_{\mathrm{DQ4}} = a_{2\mathrm{d}} \cos(2\Delta_\mathrm{d}\tau - 2\psi_\mathrm{d}). (4)$$

Note that both the amplitude $a_{2\mathrm{d}}$ and the parasitic phase shift $\psi_\mathrm{d}$ are functions of the common- and differential-mode detunings $(\Delta_\mathrm{c}, \Delta_\mathrm{d})$.

To explicitly clarify the signal dependence on $\Delta_\mathrm{c}$ and $\Delta_\mathrm{d}$, we evaluate analytical solutions for two special cases. Let $\Omega_1$ and $\Omega_2$ be the SQ Rabi angular frequencies for the $|0\rangle \leftrightarrow |+1\rangle$ and $|0\rangle \leftrightarrow |-1\rangle$ transitions, respectively. Under the condition of equal Rabi frequencies (i.e., $\Omega_1 = \Omega_2 \equiv \Omega_\mathrm{SQ} \equiv \Omega_\mathrm{DQ}/\sqrt{2}$), the dependence on $\Delta_\mathrm{d}$ in the absence of common-mode detuning ($\Delta_\mathrm{c} = 0$) is derived as:

$$S_{\mathrm{DQ4}}(\Delta_\mathrm{c} = 0) = A_{\mathrm{DQ4}} \cos(2\Delta_\mathrm{d}\tau + \Phi), (5)$$

$$A_{\mathrm{DQ4}} = \frac{1}{2}\left[\frac{\Omega_\mathrm{DQ}}{\Omega'_{\mathrm{DQ,d}}} \sin\left(\frac{\Omega'_{\mathrm{DQ,d}} t_\mathrm{p}}{2}\right)\right]^4 \times \left[1 + \left(\frac{2\Delta_\mathrm{d}}{\Omega'_{\mathrm{DQ,d}}}\right)^2 \sin^2\left(\frac{\Omega'_{\mathrm{DQ,d}} t_\mathrm{p}}{2}\right)\right]^2$$

$$\Phi = 4\tan^{-1}\left(\frac{2\Delta_\mathrm{d}}{\Omega'_{\mathrm{DQ,d}}} \tan\left(\frac{\Omega'_{\mathrm{DQ,d}} t_\mathrm{p}}{4}\right)\right)$$

where, the effective DQ Rabi frequency is $\Omega'_{\mathrm{DQ,d}} \equiv \sqrt{\Omega_\mathrm{DQ}^2 + (2\Delta_\mathrm{d})^2}$.

Eq. (5) describes the fundamental magnetic field response of DQ 4-Ramsey sensing. The phase term $\Phi$ accounts for the phase shift induced by the detuning $\Delta_\mathrm{d}$ during the MW pulse duration. In the regime where $2\Delta_\mathrm{d} \ll \Omega_\mathrm{DQ}$ under an optimal DQ $\pi$ pulse, this function can be approximated as a $\cos(2\Delta_\mathrm{d}\tau)$ interference profile. Note that there is a degree of freedom in selecting the relative phase of the second MW pulses, as discussed in Ref. [11]. For instance, shifting the relative phase of the second pulses by $\pi/2$ converts the response into a $\sin(2\Delta_\mathrm{d}\tau)$ profile. In this study, we set the phase to operate with a cosine profile, as illustrated in Fig. 1(b). To maximize the magnetic field sensitivity, a fixed detuning offset is introduced to operate the sensor at the zero-crossing point of this periodic response, where the signal slope with respect to $\Delta_\mathrm{d}$ is maximized.

Conversely, the dependence on $\Delta_\mathrm{c}$ in the absence of differential detuning ($\Delta_\mathrm{d} = 0$) is given by:

$$S_{DQ4}(\Delta_\mathrm{d} = 0) = \frac{1}{2}\left[\frac{\Omega_\mathrm{DQ}}{\Omega'_{\mathrm{DQ,c}}} \sin\left(\frac{\Omega'_{\mathrm{DQ,c}} t_\mathrm{p}}{2}\right)\right]^4, (6)$$

where the effective DQ Rabi frequency is $\Omega'_{\mathrm{DQ,c}} \equiv \sqrt{\Omega_\mathrm{DQ}^2 + \Delta_\mathrm{c}^2}$. In the regime where $\Delta_\mathrm{c} \ll \Omega_\mathrm{DQ}$ under a DQ $\pi$ pulse, this function can be approximated as a $\exp\left[-2\left(\Delta_\mathrm{c}/\Omega_\mathrm{DQ}\right)^2\right]$ Gaussian profile. Fig. 1(c) illustrates the resulting suppression of the signal amplitude with increasing $|\Delta_\mathrm{c}|$. Temperature fluctuations shift the ZFS, thereby directly degrading the DQ 4-Ramsey signal amplitude. This amplitude suppression reduces the amplitude of $\Delta_\mathrm{d}$ interference fringes, as clearly observed in Fig. 1(b). Consequently, the suppression of the signal amplitude by $\Delta_\mathrm{c}$ reduces the magnetic signal slope, thereby degrading the magnetic field sensitivity. To enable high-sensitivity and high-stability magnetic field sensing, both the differential- and common-mode drifts must be addressed.

However, conventional DQ 4-Ramsey magnetometry faces a fundamental dilemma. The sequence is operated at the zero-crossing of the magnetic-field response to maximize $\left|\frac{\partial S_{\mathrm{DQ4}}}{\partial \Delta_\mathrm{d}}\right|$ (Fig. 1(b)). At this operating point, the first-order response to common-mode detuning vanishes $\left(\frac{\partial S_{\mathrm{DQ4}}}{\partial \Delta_\mathrm{c}} = 0\right)$, preventing direct detection of common-mode frequency drift. Thus, optimizing the magnetic field sensitivity inherently suppresses the available first-order error signal for frequency tracking. This motivates us to develop a method that can simultaneously track common- and differential-mode frequency drifts without compromising the intrinsic magnetic field sensitivity.

TABLE I. MW phase combinations and the signs of corresponding signal components in the DQ-1 Ramsey signal.

| | Phase alterations | | | | Corresponding signs | | | |
|---|---|---|---|---|---|---|---|---|
| Label | $\phi_1$ | $\phi_2$ | $\phi_c$ | $\phi_d$ | DC | DQ($+1$: $-1$) | SQ($+1$: 0) | SQ($-1$: 0) |
| A | 0 | 0 | 0 | 0 | + | + | + | + |
| B | $\pi$ | 0 | $\pi/2$ | $\pi/2$ | + | − | − | + |
| C | $\pi$ | $\pi$ | $\pi$ | 0 | + | + | − | − |
| D | 0 | $\pi$ | $\pi/2$ | $-\pi/2$ | + | − | + | − |

TABLE II. Representation of DQ 4-Ramsey signals using four combinations of DQ 1-Ramsey signals.

| Label | DC | DQ($+1 \leftrightarrow -1$) | SQ($+1 \leftrightarrow 0$) | SQ($-1 \leftrightarrow 0$) |
|---|---|---|---|---|
| $(A+B+C+D)/4$ | $c_0$ | 0 | 0 | 0 |
| $(A-B+C-D)/4$ | 0 | $a_{2d}\cos(2\Delta_d\tau - 2\psi_d)$ | 0 | 0 |
| $(A-B-C+D)/4$ | 0 | 0 | $a_1\cos(\Delta_1\tau - \psi_1)$ | 0 |
| $(A+B-C-D)/4$ | 0 | 0 | 0 | $a_2\cos(\Delta_2\tau - \psi_2)$ |

### B. Dual Feedback Sensing Scheme

To compensate for operating-point deviations induced by environmental magnetic-field and temperature fluctuations without compromising magnetic field sensitivity, we implement a dual-feedback sensing scheme integrated into the DQ 4-Ramsey architecture, as illustrated in Fig. 2. Fig. 2(a) shows the standard DQ 4-Ramsey sequence, which consists of four consecutive phase-cycling steps labeled A, B, C, and D. Demodulation at $f_0$ yields the DQ 4-Ramsey signal, which exhibits a coherence-dependent cosine-type response to the differential-mode detuning $\Delta_d$, as shown in Fig. 2(b). In contrast, the DC component of the fluorescence signal contains the population term $c_0$, which corresponds to the A+B+C+D variant of the DQ 4-Ramsey signal. $c_0$ is independent of the free-precession time and provides a response to the common-mode detuning $\Delta_c$. The DQ and $c_0$ responses are therefore employed as error signals for magnetic-field and temperature stabilization, respectively.

However, both responses exhibit zero slope at zero detuning (Fig. 2(b)). We therefore introduce synchronized common- and differential-mode frequency modulations to obtain steep-slope error signals at the operating point. As illustrated in Fig. 2(c), square-wave frequency deviations are synchronized with the repeated DQ 4-Ramsey sequence and remain constant over each four-step block. The differential- and common-mode modulation frequencies are set to $f_0/4$ and $f_0/8$, respectively. These frequencies can be adjusted based on the required signal spectral separation and detection bandwidth.

Fig. 2(d) illustrates the resulting frequency-domain representation of the fluorescence signal. Without the modulation, the $c_0$ and DQ responses appear at DC and $f_0$, respectively, as indicated by the orange and blue dashed lines. With the synchronized modulation, the $c_0$ response is shifted from DC to $f_0/4$, while the DQ response is shifted from $f_0$ to the sidebands at $f_0 \pm f_0/8$. These components are spectrally separated, enabling independent extraction of the common- and differential-mode error signals. Demodulation at $f_0/4$ yields a derivative-like response for $c_0$, while demodulation at $f_0 \pm f_0/8$ yields a sine-type DQ response. The exact profiles of these responses depend on both the modulation depths $f_{dev,c}$ and $f_{dev,d}$. By optimizing the modulation depths, both responses provide a high sensitivity around zero detuning, enabling the extraction of linear error signals.

The extracted error signals are used to independently stabilize temperature- and magnetic-field-induced frequency shifts, as shown in Fig. 2(e). The common-mode error signal obtained from $c_0$ is fed back to the local oscillator (LO) through frequency modulation (FM), thereby adjusting the MW frequency to compensate for common-mode resonance-frequency shifts. In parallel, the differential-mode DQ error signal is fed back to a loop coil to compensate for magnetic-field fluctuations through the corresponding Zeeman shift.

For magnetic-field signals within the differential-mode feedback bandwidth, the feedback control signal itself can be used for detection, whereas signals outside this bandwidth can be detected directly from the fluorescence signal. Furthermore, the common-mode feedback signal provides a direct readout of the temperature-induced frequency shift. By dividing this shift by the calibrated temperature coefficient ($-74$ kHz/K [12]), the actual temperature variation in the diamond can be determined. Thus, this dual-feedback architecture enables simultaneous stabilization and multimodal sensing of magnetic-field and temperature variations.

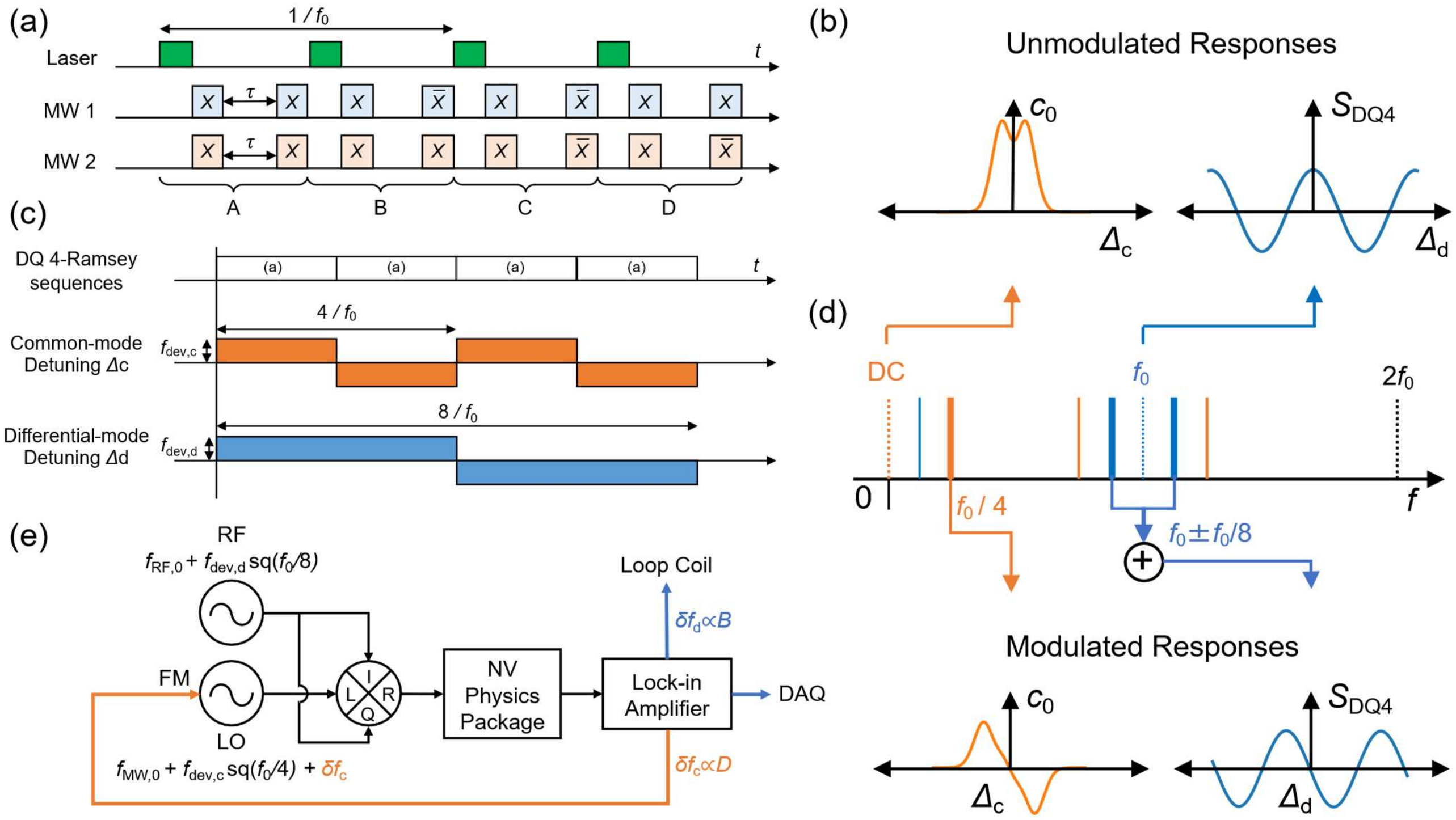

FIG. 2. Overview of the dual-feedback architecture. (a) Standard DQ 4-Ramsey sequence composed of four consecutive phase-cycling steps labeled A, B, C, and D. (b) Unmodulated DQ 4-Ramsey responses, showing the population response $c_0$ as a function of $\Delta_c$ and the cosine-type DQ response as a function of $\Delta_d$. (c) Synchronized modulation scheme employing square-wave frequency deviations that remain constant within each four-step block. (d) Frequency-domain representation of the fluorescence signal and corresponding demodulated responses: a dispersive response at $f_0/4$ and a sine-type magnetic response at $f_0 \pm f_0/8$. (e) Schematic representation of the microwave generation and feedback setup. The $\Delta_c$ error signal is fed back to the LO via FM (orange line) for active compensation of common-mode frequency shifts, while the $\Delta_d$ error signal is fed back to the loop coil for the stabilization of differential-mode frequency shifts and magnetic-field detection.

## III. Demonstration of Dual-Feedback Sensing Scheme in DQ 4-Ramsey Magnetometry

### A. Experimental Methods

The diamond was synthesized using a high-pressure and high-temperature method with a $^{12}C$ isotopically enriched carbon source ($^{13}C$=50 ppm) [25]. The synthesized diamond was cut parallel to the (111) plane with a thickness of 0.31 mm. Electron-beam irradiation at a fluence of $1 \times 10^{17}$ cm$^{-2}$ was carried out to form $NV^-$ centers. The concentration of P1 and $NV^-$ centers were estimated by electron paramagnetic resonance as 0.4 ppm and 0.07 ppm, respectively. To employ the light-trapping diamond waveguide technique [26,13,14], the diamond was further shaped into a $1.6 \times 1.6$ mm$^2$ square featuring a 45° angled facet of 0.31 mm at one corner.

A bias magnetic field of 2.3 mT was applied along the [111] direction using a pair of magnet arrays consisting of 72 Sm-Co magnets [13,27,14]. The dephasing time $T_2^*$ was measured to be 5.0 μs for the SQ basis and 10.4 μs for the DQ basis.

Fig. 3(a) illustrates the optical, MW, and RF experimental setup. A 532-nm laser beam was delivered through an optical fiber to the sensor head in a three-layer magnetic shield box. An optical power of 20–200 mW was focused onto the diamond using a convex lens with a focal length of 150 mm. The acousto-optic modulator (AOM) was stabilized by proportional-integral (PI) control implemented on a Red Pitaya Field-Programmable Gate Array (FPGA) platform [28]. This feedback stabilization was used to suppress $1/f$ intensity noise of the laser. Balanced detection was implemented following the method described in Ref. [14].

We implemented a MW chain with a mixing configuration composed of MW and RF components [24,29]. To enhance the spin contrast, six-tone MW control was employed to simultaneously drive all three $^{14}N$ hyperfine states for both the $m_s = 0 \leftrightarrow \pm 1$ transitions. The MW pulses were shaped into a Gaussian profile, and the MW power was limited to avoid off-resonant driving of adjacent hyperfine states [13,30]. To obtain the desired dispersive response, we equalized the DQ

Rabi frequencies for all the hyperfine states. Further details regarding the mixing configuration are provided in Appendix B. The NV spins were driven using a printed-circuit-board (PCB) antenna [27].

We set the modulation frequency to $f_0 = 20\ \text{kHz}$, which comprised a laser irradiation time of 13 μs and a free precession time of $\tau = 7.1\ \mu\text{s}$. The signal generator (SG) was frequency-modulated at a modulation frequency of $f_0/4 = 5\ \text{kHz}$ with a deviation frequency of $f_{\text{dev,c}} = 180\ \text{kHz}$ (higher robust mode), or $f_{\text{dev,c}} = 40\ \text{kHz}$ (higher magnetic sensitivity mode). The arbitrary waveform generator (AWG) was programmed to generate a modulation frequency of $f_0/8 = 2.5\ \text{kHz}$ with a deviation frequency of $f_{\text{dev,d}} = 50\ \text{kHz}$. The differential-mode signal was first demodulated at 20 kHz, followed by further demodulation at 2.5 kHz. The common-mode signal was demodulated at 5 kHz.

Fig. 3(b) shows a close-up schematic of the sensor head. To demonstrate active common-mode shift tracking, a thermoelectric cooler (TEC) was employed as a heating element. The TEC was thermally coupled to the sensor head via a heat pipe to avoid electromagnetic disturbances to the MW driving caused by direct antenna contact. The heat pipe was attached to the hot side of the TEC using thermally conductive double-sided tape, and the TEC driving current was regulated by a TEC current controller. The PI feedback signal for differential-mode detuning compensation was applied to a home-built 10-turn coil. The test magnetic-field amplitude was calibrated by comparing the experimentally detected value with calculation using the Biot-Savart law.

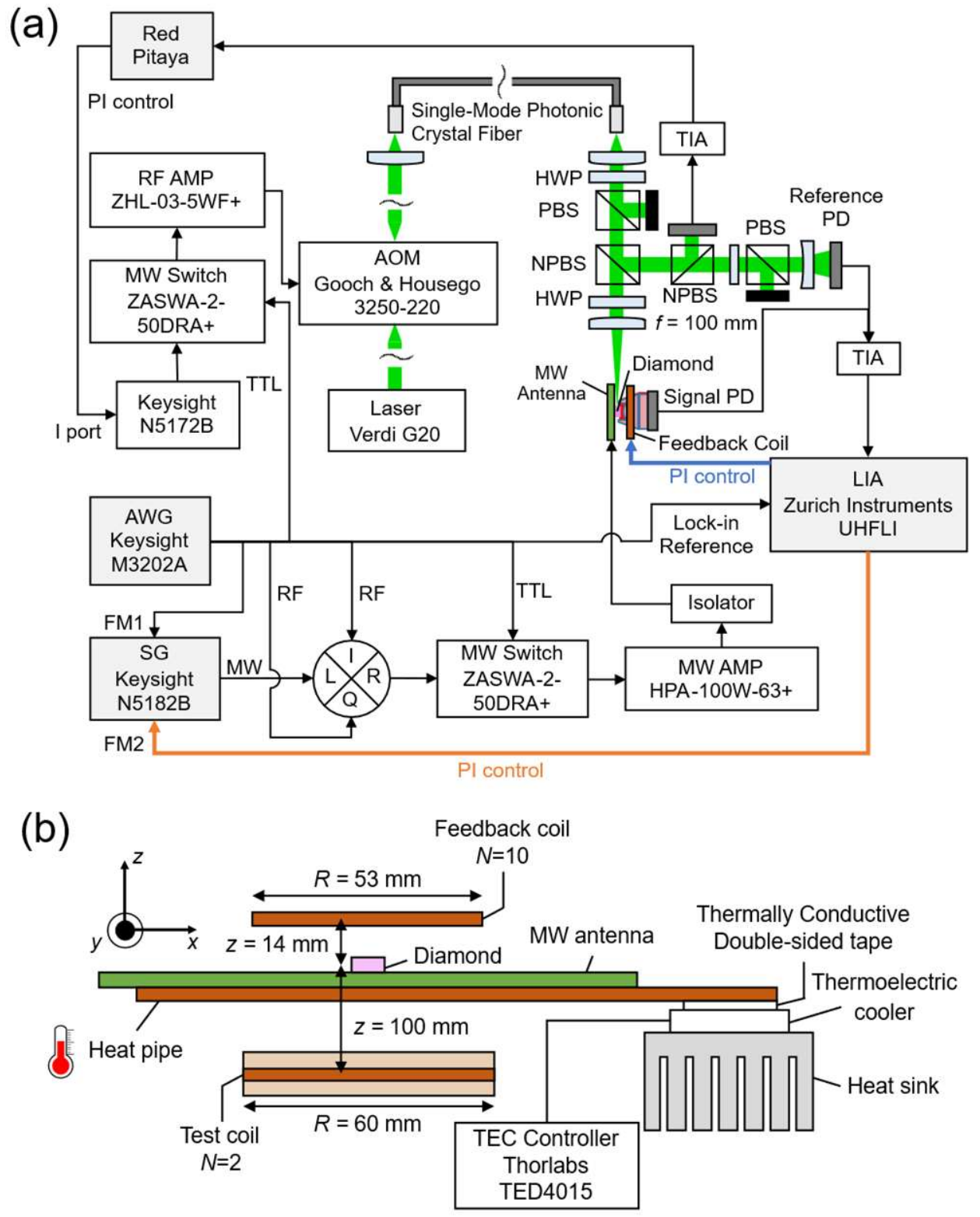

FIG. 3. Experimental setup. (a) Optical, MW, and RF setup. The laser intensity is stabilized via an AOM using an FPGA board. A photonic crystal fiber delivers the laser light to the sensor head. (b) Close-up sensor head schematic integrated with a heat pipe and coils. The magnetic field to be sensed is aligned along the z axis, $[111]_{\text{diamond}}$. The heat pipe transfers the heat generated by the TEC. A TEC controller applies a DC current (< 1A) to the TEC.

### B. Optimization of Modulation Depth

While the analytical formalism in Section II was derived assuming rectangular MW pulses, Gaussian-shaped MW pulses were utilized in the actual sensing experiments. Because each NV center contains $^{14}$N isotope, off-resonant driving of adjacent hyperfine states can disturb the desirable spin dynamics. To suppress such crosstalk while maintaining sufficient signal contrast, the three hyperfine states of $^{14}$N were simultaneously driven with a relatively low Rabi frequency ($\Omega_{\text{DQ}}/2\pi \ll 2.16\ \text{MHz}$) using Gaussian-shaped MW pulses [13,30]. In the regime where the detuning is small compared to the Rabi frequency, the analytical solution derived for rectangular pulses accurately describes the response of Gaussian pulses, allowing us to analytically predict the optimal modulation depth.

Fig. 4(a) shows the comparison between the experimental and theoretical differential-mode response of the DQ term as functions of the deviation frequency $f_{\text{dev,d}}$ and detuning $\Delta_{\text{d}}$. The experimental result obtained with Gaussian pulses can be related to the analytical model for rectangular MW pulses through a correction to the free-precession time:

$$\tau_{\mathrm{rect}} \approx \tau_{\mathrm{gauss}} - \frac{4}{\Omega_{\mathrm{DQ}}}. (7)$$

This relation is derived from Eq. (5) in the limit of $2\Delta_{\mathrm{d}} \ll \Omega_{\mathrm{DQ}}$, while a MW pulse duration $t_{\mathrm{p}}$ for the rectangular pulse is chosen to realize a DQ $\pi$ pulse at $\Delta_{\mathrm{d}} = 0$. Here, $\tau_{\mathrm{rect}}$ is defined as the interval between the end of the first MW pulse and the beginning of the second MW pulse, as illustrated in Fig. 2(a). In contrast, $\tau_{\mathrm{gauss}}$ is defined as the interval between the centers of the two Gaussian pulses, as depicted in Fig. 7 in Appendix B. For instance, with $\Omega_{\mathrm{DQ}}/2\pi = 300$ kHz and $\tau_{\mathrm{gauss}} = 7.17\ \mu s$, Eq. (7) yields $\tau_{\mathrm{rect}} \approx 5.04\mu\mathrm{s}$. Using this relation, the analytical model reproduces the experimental $\Delta_{\mathrm{d}}$ dependence.

Since the phase shift $\Phi$ in Eq. (5) was negligible in the Gaussian-pulse experiment, the maximum magnetic-field slope should be obtained when the differential-mode deviation frequency satisfies

$$f_{\mathrm{dev,d,opt}} \approx \frac{2n-1}{8\tau_{\mathrm{gauss}}}, (n = 1,2,3 \ldots). (8)$$

As indicated by the white dashed lines in Fig. 4(a), the response exhibits the highest contrast at optimal deviation frequencies of $f_{\mathrm{dev,d,opt}} \approx 17$ kHz and 52 kHz, yielding comparable maximum slopes. Furthermore, as shown in Fig. 4(b), introducing this differential-mode modulation converts the cosine profile into a sine profile while maintaining its amplitude. This results in a minor reduction of only 7% in the maximum slope compared with the unmodulated case.

Fig. 4(c) shows the comparison between the experimental and numerical common-mode responses of the $c_0$ term as functions of the deviation frequency $f_{\mathrm{dev,c}}$ and detuning $\Delta_{\mathrm{c}}$. The specific solution for $c_0$ at $\Delta_{\mathrm{d}} = 0$, assuming rectangular MW pulses, is expressed as:

$$c_0(\Delta_{\mathrm{d}} = 0) = 1 - 2\left[\frac{\Omega_{\mathrm{DQ}}}{\Omega'_{\mathrm{DQ,c}}}\sin\left(\frac{\Omega'_{\mathrm{DQ,c}}t_{\mathrm{p}}}{2}\right)\right]^2 + \frac{3}{2}\left[\frac{\Omega_{\mathrm{DQ}}}{\Omega'_{\mathrm{DQ,c}}}\sin\left(\frac{\Omega'_{\mathrm{DQ,c}}t_{\mathrm{p}}}{2}\right)\right]^4. (9)$$

As illustrated in the left panel of Fig. 2(b), this profile exhibits a slight central dip flanked by two peaks. However, the specific solution cannot be directly used to determine the optimal deviation frequency under the experimental conditions, primarily because it assumes rectangular MW pulses and neglects relaxation effects. We therefore evaluated the optimal deviation frequency under the experimental conditions and validated it alongside numerical simulations based on the Lindblad master equation, as shown in Fig. 4(c). Fig. 4(d) shows one-dimensional profiles extracted along the white dashed lines in Fig. 4(c), demonstrating general agreement between the numerical (dashed) and experimental (solid) results. Modulation with $f_{\mathrm{dev,c}} = 180$ kHz provides a broad linear range (orange), whereas the linearity near zero detuning is minimal at $f_{\mathrm{dev,c}} = 40$ kHz (blue). Since the magnetic-field slope decays as the deviation frequency $f_{\mathrm{dev,c}}$ increases, we adopted $f_{\mathrm{dev,c}} = 180$ kHz for a highly robust mode and $f_{\mathrm{dev,c}} = 40$ kHz for the highly sensitive mode in Secs. III. C and D, respectively.

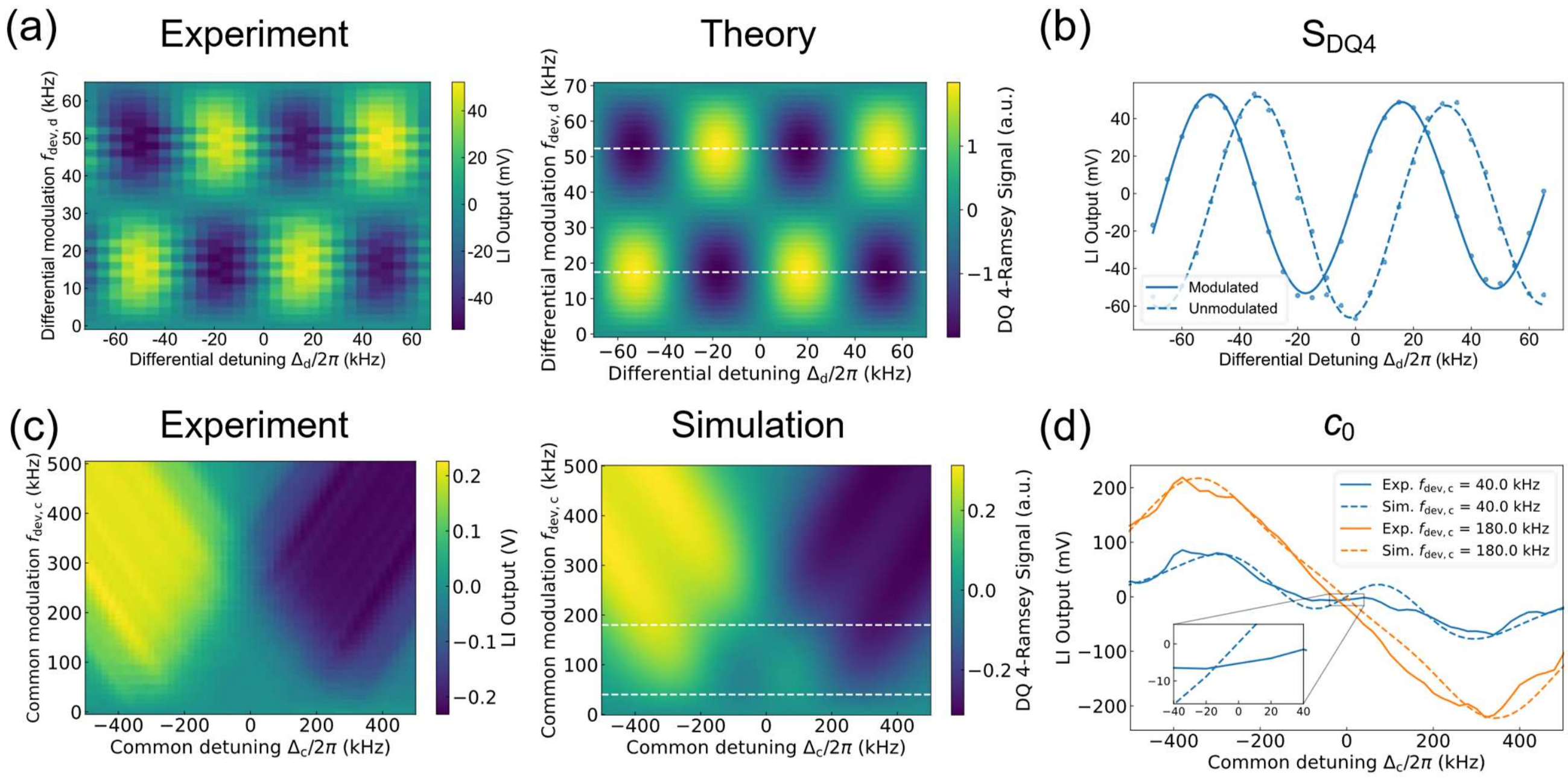

FIG. 4. (a) Experimental and numerically calculated differential-mode responses as functions of $f_{\mathrm{dev,d}}$ and $\Delta_{\mathrm{d}}$. White dashed lines indicate the optimal deviation frequencies of $f_{\mathrm{dev,d,opt}} \approx 17$ kHz and 52 kHz. (b) Differential-mode DQ responses with and without modulation. Modulation converts the cosine profile into a sine profile while maintaining amplitude. (c) Experimental and numerically calculated common-mode responses as functions of $f_{\mathrm{dev,c}}$ and $\Delta_{\mathrm{c}}$. White dashed lines indicate the deviation frequencies selected for sensing experiments. (d) Corresponding modulated $c_0$ responses. Solid and dashed lines represent experimental and numerical results, respectively. Modulation with $f_{\mathrm{dev,c}} = 180$ kHz provides a broad linear range, whereas $f_{\mathrm{dev,c}} = 40$ kHz yields minimal linearity. The inset shows the non-zero slope at the zero-crossing point.

### C. Demonstration of Active Compensation for Temperature Variations

To demonstrate the active tracking architecture, we applied a temperature perturbation to the diamond sensor using the TEC, as shown in Fig. 3(b). A burst of test magnetic field was repeatedly applied with its amplitude detectable by single-shot measurements. Fig. 5 demonstrates the performance of the dual-feedback sensing scheme, compared with the conventional open-loop scheme. Fig. 5(a) shows the time traces of the measured signals. The inset displays a single-period (500 ms) time window, in which the applied magnetic field (40 Hz, 4 cycles) is clearly resolved. The time axis was aligned with the onset of the thermal perturbation (dashed black line) driven by applying a current of $-0.5$ A to the TEC. In the conventional open-loop scheme, the 40-Hz oscillation amplitude gradually changed as the diamond temperature drifted. In contrast, with the feedback, the oscillation amplitude remained stable despite the temperature perturbation.

Fig. 5(b) shows the estimated amplitude of the applied magnetic field obtained by fitting the measured signals in Fig. 5(a). The fitted amplitude was converted to the magnetic-field amplitude using the differential-mode slope determined prior to the measurement. The two traces initially agreed well; however, the open-loop signal gradually decreased, whereas the signal obtained with PI feedback remained nearly constant throughout the measurement. Fig. 5(c) shows the LO frequency shift applied by the PI controller, which can be converted to the temperature variations in diamond using a ZFS temperature coefficient of $-74$ kHz/K[12]. The open-loop signal became completely undetectable at a temperature increase of 2.5 K, whereas the PI feedback sustained the magnetic field sensitivity up to 2.8 K, which was reached at the end of the measurement.

These results demonstrate that the active tracking architecture compensates for thermal frequency shifts while maintaining magnetic field sensitivity, thereby enabling simultaneous multimodal sensing in Ramsey magnetometry.

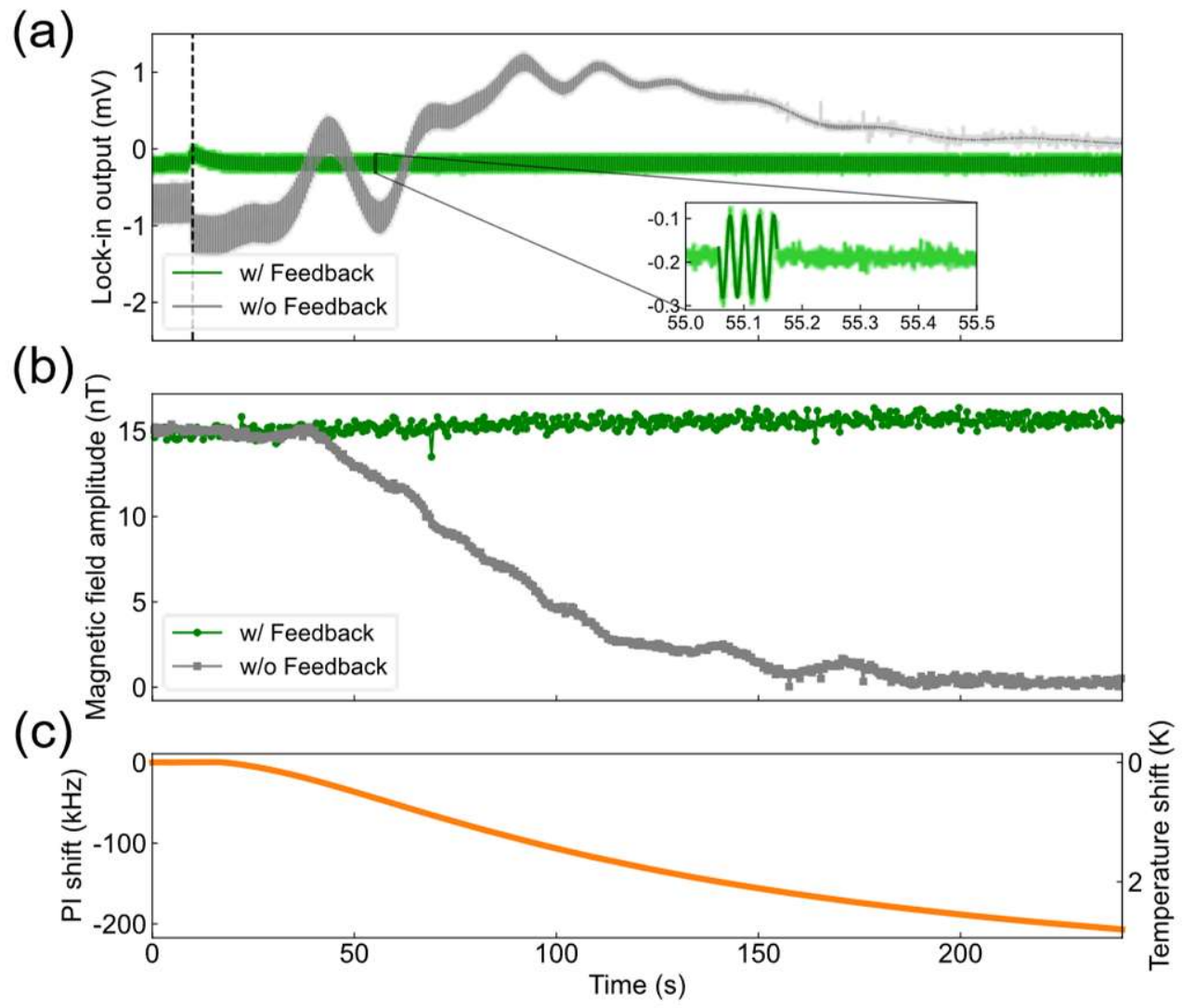

FIG 5. Demonstration of active feedback for temperature shift. (a) Time traces of the DQ 4-Ramsey signals with (green) and without (gray) feedback. (b) Extracted magnetic field amplitude obtained by fitting the measured signal. (c) Frequency shift applied by the PI controller, corresponding to the temperature-induced frequency shift of the diamond.

### D. Highly Stable and Sensitive Measurements

To evaluate the long-term stability of the sensor, we measured the Allan deviation with and without feedback under ambient conditions and without intentional TEC heating, as shown in Fig. 6(a). The gray, orange, blue, and green curves correspond to the PI-off condition, common-mode feedback, differential-mode feedback, and dual-feedback, respectively. The feedback bandwidths and PI gains for the common- and differential-mode loops were independently configured. A bandwidth of 1 Hz was applied to both modes in these measurements. In all four cases, the Allan deviation initially decreased following the $\tau^{-1/2}$ dependence expected for white noise, but increased at approximately $\tau = 0.4$ s following the $\tau^{1/2}$ dependence. While the deviation continued to increase under both the PI-off and common-mode feedback conditions, it began to decrease again for $\tau > 10$ s with the differential-mode feedback and dual-feedback. This indicates that common-mode thermal fluctuation was not a dominant noise source at this time scale. This decrease in the Allan deviation is attributed to the suppression of magnetic $1/f$ noise by the differential-mode feedback. This is consistent with the magnetic noise reduction below approximately 0.1 Hz observed in the double-sided amplitude spectrum density (ASD) shown in the inset panel of Fig. 6(a).

We evaluated the magnetic field sensitivity, and the achievable magnetic noise floor with dual-feedback. Fig. 6(b) shows the positive-frequency region of the double-sided ASD of the magnetic field. The spectrum with the dual feedback (blue) gives an estimate of the magnetic field sensitivity with dual-feedback. We estimated a magnetic field sensitivity of $7.0\ \mathrm{pT}/\sqrt{\mathrm{Hz}}$ in the frequency range of $5 - 100$ Hz by using a moving-average fit. The blue dotted line indicates the photon-shot-noise (PSN) limited sensitivity of $1.5\ \mathrm{pT}/\sqrt{\mathrm{Hz}}$, calculated using the following equation [14]:

$$\eta_{\mathrm{Ramsey}}^{\mathrm{ensemble,shot}} = \frac{\sqrt{2qI}}{\gamma_e/2\pi \left|\frac{dI}{d\Delta_d/2\pi}\right|_{\mathrm{exp}}}, \quad (10)$$

where $q = 1.6 \times 10^{-1}$ C is the elementary charge, $\left|\frac{dI}{d\Delta_d/2\pi}\right|_{\mathrm{exp}} = 521$ pA/Hz is the experimentally obtained maximum slope, and $I = 1.5$ mA is the time-averaged photocurrent. The discrepancy between the measured sensitivity and the PSN limit is likely to arise from optical noise. By maintaining the magnetic field sensitivity for an extended duration of approximately 4 hours with dual feedback, we achieved a noise floor of $64\ \mathrm{fT}/\sqrt{\mathrm{Hz}}$, as indicated by the green spectrum. This reduction closely follows the $1/\sqrt{N}$ scaling expected for uncorrelated white noise, where $N$ is the number of averages. The green dotted line represents the PSN-limited sensitivity divided by $\sqrt{N}$ ($N = 14449$), corresponding to $12\ \mathrm{fT}/\sqrt{\mathrm{Hz}}$.

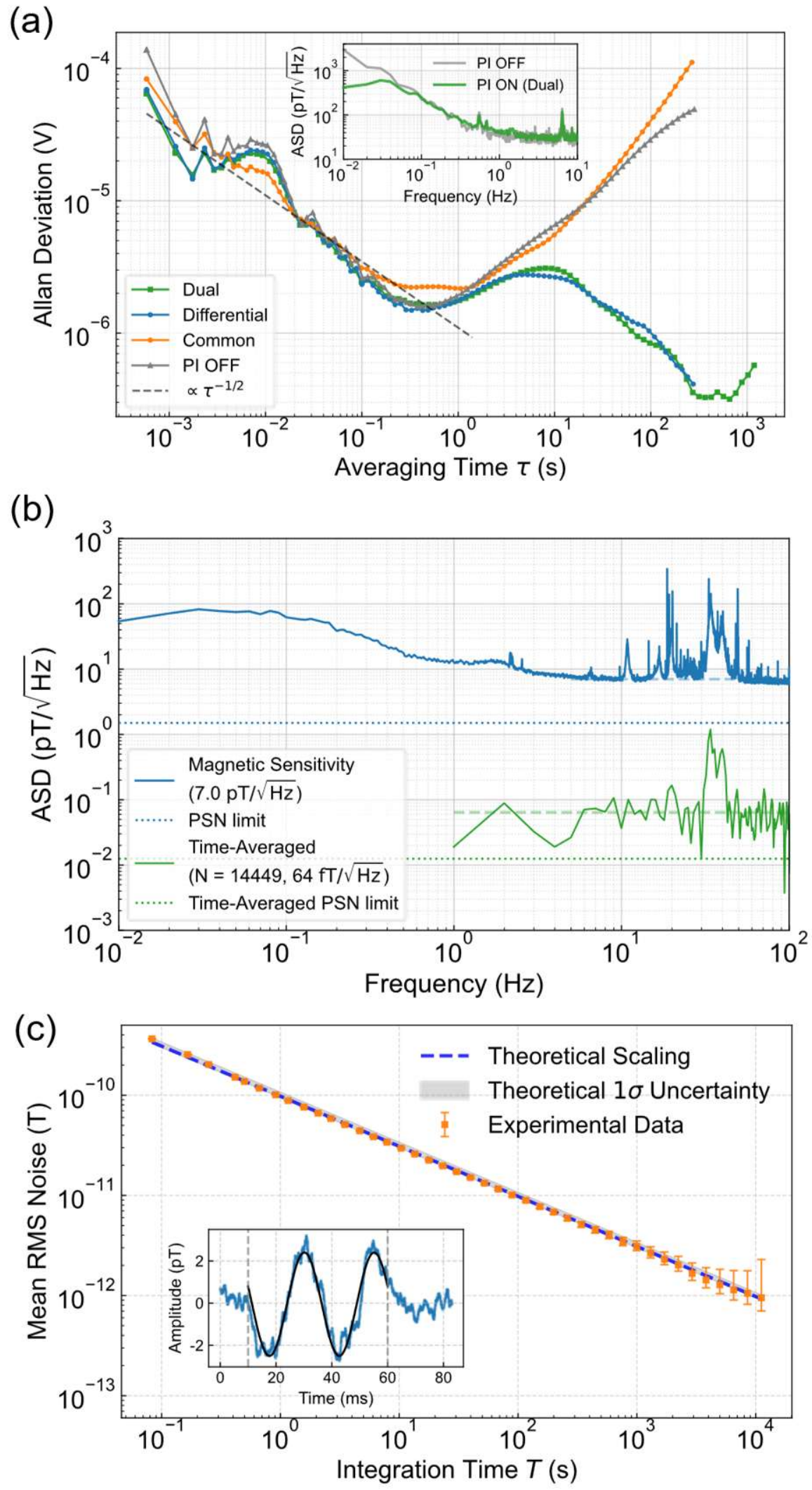

FIG. 6. Long-term operation. (a) Allan deviation of the DQ 4-Ramsey signals under varying feedback conditions: open-loop (gray), common-mode feedback (orange), differential-mode feedback (blue), and dual-feedback (green). Inset: Corresponding ASDs. (b) Double-sided ASD of the magnetic field with feedback (blue). The green spectrum represents the ASD of the signal averaged over approximately 4 h, using 1-s measurement epochs. The fitted lines indicate noise floors of 7.0 pT/$\sqrt{\text{Hz}}$ without averaging, and 64 fT/$\sqrt{\text{Hz}}$ with averaging, respectively. (c) Magnetic field noise scaling. Data points represent the mean RMS noise of the waveforms averaged over the integration time $T$, using non-overlapping 83-ms windows. Error bars show the $1\sigma$ confidence intervals. The blue dashed line and the gray shaded region indicate the theoretical $T^{-1/2}$ white-noise scaling and its $1\sigma$ uncertainty, respectively. The data follows this scaling up to 4 hours. The inset panel shows the averaged waveform of a detected weak magnetic-field signal of approximately 2.5 pT.

To characterize the averaging behavior over an extended integration time, we measured a weak magnetic field via signal accumulation. A test magnetic field was applied using the test coil illustrated in Fig. 3(b). A current of 2.5 μA was supplied to the coil, producing a calibrated magnetic-field amplitude of 2.5 pT based on the Biot-Savart law. The magnetic field was applied as a burst signal generated by a function generator (FG), consisting of a two-cycle, 40-Hz sinusoidal waveform repeated every 85 ms. The measured Ramsey signal was processed using a fourth-order low-pass filter with a cutoff frequency of 200 Hz, and accumulated every 83 ms using a DAQ system triggered by the FG. Concurrently, the signal was sampled at a rate of 214 Hz to analyze the RMS noise scaling. The mean RMS noise of the averaged waveforms within the 83-ms time window was evaluated as a function of the total integration time $T$, as shown in Fig. 6(c). Here, $T$ is defined as the total accumulated time of the windows ($T = N\Delta T$), where $\Delta T = 83$ ms is the time window, and $N$ is the number of averages. As indicated by the orange data points, the RMS noise decreased consistently with the $T^{-1/2}$ scaling expected for uncorrelated noise. The RMS noise scaling was calculated using the following equation:

$$\sigma = \eta\sqrt{f_s/N}, (11)$$

where $\eta = 23.1$ pT/$\sqrt{\text{Hz}}$ is the magnetic field sensitivity for this measurement, $f_s = 214$ Hz is the sampling rate. As indicated by the blue dashed line, this theoretical scaling is in agreement with the measured data.

After $N = 1.6 \times 10^6$ averages, corresponding to an integration time of approximately 38 hours, the weak magnetic field signal became clearly visible, as shown in the inset of Fig. 6(c). In the averaged waveform, while the uncorrelated white noise is substantially suppressed, residual noise synchronized with the trigger timing remains the dominant noise source at this integration timescale. Nevertheless, the fitted amplitude of 2.5 pT is in good agreement with the expected value, exhibiting neither significant waveform distortion nor amplitude attenuation. These results demonstrate that the dual-feedback scheme effectively maintains both the intrinsic magnetic field sensitivity and the optimal operating point over a prolonged duration, enabling the reliable detection of signals substantially below the single-shot ASD noise level.

## IV. Discussion

In this work, we proposed a dual-feedback scheme to dynamically compensate for both common- and differential-mode detuning. By maintaining a magnetic field sensitivity of 7.0 pT/$\sqrt{\text{Hz}}$ in the frequency range of $5 - 100$ Hz, we achieved a magnetic noise floor of 64 fT/$\sqrt{\text{Hz}}$ over 4 hours of integration, which reaches the required detection threshold for MEG (typically $\approx 100$ fT at $< 100$ Hz). If the intrinsic magnetic field sensitivity can be improved to the PSN limit (1.5 pT/$\sqrt{\text{Hz}}$) as shown in Fig. 6(b), the required integration time to reach the 64 fT/$\sqrt{\text{Hz}}$ noise floor would be reduced to approximately 10 minutes. Furthermore, since biomagnetic fields typically decay as $r^{-2}$ or faster, diamond quantum sensors, with a significantly shorter sensor-to-sample distance (e.g., 2 mm [14]) compared to OPMs (6 mm [31,32]) or SQUIDs (2∼3 cm [32]), offer a distinct spatial advantage for MEG measurements.

However, high temporal resolution is essential for practical biomedical applications. Even state-of-the-art OPM-MEG studies, despite possessing an ultrahigh intrinsic sensitivity (∼15 fT/$\sqrt{\text{Hz}}$), rely on signal averaging across dozens of repeated trials (e.g., 50 trials) to characterize task-induced neural responses [8]. Consequently, compensating for the current sensitivity limitations of NV centers solely through extended integration times is impractical for capturing real-world neural activity. To attain the requisite temporal resolution, further improvements in the intrinsic magnetic field sensitivity are indispensable.

Fundamentally, the sensitivity is improved by increasing the photocurrent, enhancing the readout contrast, or prolonging the spin dephasing time. However, these sensitivity-enhancement strategies introduce a fundamental trade-off between sensitivity and robustness. For instance, increasing $T_2^*$ allows for a longer free-precession time $\tau$, thereby improving sensitivity, yet, it simultaneously narrows the linear operating range with respect to differential-mode detuning, scaling approximately as $|\Delta_\mathrm{d}| \leq 1/(4\tau)$. Thus, sensors optimized for long coherence times become susceptible to magnetic-field drifts, such as residual geomagnetic fields. Additionally, a similar challenge arises from thermally induced frequency drifts. Approaches such as increased optical excitation power and MW power [17], as well as RF-based techniques such as spin-bath driving (SBD) [13,33], inevitably introduce additional thermal loads that shift the ZFS and destabilize the resonance condition. Moreover, highly advantageous close-proximity operation of diamond quantum sensors makes them more susceptible to heat originating from living samples. The dual-feedback scheme proposed here can effectively address these limitations, proving particularly advantageous for such high-sensitivity configurations. Because our active tracking architecture is fully compatible with the standard DQ 4-Ramsey sequence, it actively stabilizes the optimal operating point against both thermally induced shifts and environmentally driven magnetic drifts, fully preserving the benefits of the enhanced intrinsic sensitivity.

It is also worth noting that our feedback scheme has advantages over dual-resonance CW-ODMR tracking techniques [18,34,35,3]. While dual-resonance ODMR can monitor and compensate for thermal drifts, it is subject to optical and microwave power broadening, which reduces the spectral slope and consequently limits achievable magnetic field sensitivity. In addition, simultaneous driving of the two resonances can induce coherent population trapping [26,34], which further reduces the fluorescence contrast. Dual-resonance tracking also requires careful matching of the spectral slopes between the $|0\rangle \leftrightarrow |\pm 1\rangle$ transitions to properly separate common- and differential-mode responses. In contrast, our active tracking scheme is implemented within the DQ 4-Ramsey framework, where PI error signals are derived from the DQ 4-Ramsey response. By operating within this pulsed framework, our dual-feedback system inherently circumvents the limitations of CW-ODMR, thereby preserving the intrinsic magnetic field sensitivity and signal contrast without the need for precise spectral slope balancing.

Furthermore, the proposed DQ 4-Ramsey protocol with dual feedback may be applicable to challenging systems such as nanodiamond (ND) magnetometry. NDs typically exhibit short dephasing times ($T_2^* \approx 100$ ns [36]), which limits the use of standard Ramsey interferometry. SBD can extend $T_2^*$, but the strong RF driving required for dense P1 baths can induce sample heating [33,13]. Integrating dual feedback system with SBD may facilitate stable coherence-based sensing in NDs and enable multimodal measurements of temperature and magnetic fields at the nanoscale [18,34].

## ACKNOWLEDGMENTS

This work was supported by Q-LEAP Grant Numbers JPMXS0118067395, JPMXS0118068379 and JST ASPIRE Grant Number JPMJAP24C1.

## APPENDIX A: Theoretical Derivations and Hamiltonian Formalism

In the $S_z$ basis, the eigenstates $|{+1}\rangle$, $|0\rangle$, and $|{-1}\rangle$ correspond to the spin projections $m_s = +1$, $0$, and $-1$, respectively. Under the rotating-wave approximation, the effective Hamiltonians can be expressed as:

$$\widetilde{H}_0 = -\Delta_\mathrm{c} S_z^2 - \Delta_\mathrm{d} S_z$$

$$\widetilde{H}_{1+2}^{(0,0)} = \frac{\Omega}{2}\left[|B\rangle\langle 0| + |0\rangle\langle B|\right]$$

where the bright state $|B\rangle$ is defined as $|B\rangle = (\Omega_1|{+1}\rangle + \Omega_2|{-1}\rangle)/\Omega$. Here, $\Omega_1$ and $\Omega_2$ are the single-quantum Rabi frequencies for the $m_s = 0 \leftrightarrow \pm 1$ transitions as illustrated in Fig. 1(a), and $\Omega = \sqrt{\Omega_1^2 + \Omega_2^2}$ is the effective Rabi frequency. The drive Hamiltonian $\widetilde{H}_{1+2}^{(0,0)}$ thereby induces a transition between the state $|0\rangle$ and the bright state $|B\rangle$.

To evaluate the response to MW phase alterations, we set the phases of the first MW pulse to $(\phi_1, \phi_2) = (0, 0)$, while leaving the phases of the second pulse arbitrary. The propagator for the DQ 1-Ramsey sequence is given with the propagators during the two pulses and the free precession time by:

$$U_{DQ1}^{(\phi_1,\phi_2)}(\tau) = U_{0+1+2}^{(\phi_1,\phi_2)}(t_\mathrm{p})\, U_0(\tau)\, U_{0+1+2}^{(0,0)}(t_\mathrm{p}),$$

$$U_{0+1+2}^{(\phi_1,\phi_2)}(t_\mathrm{p}) = \exp\left(-i\,\widetilde{H}_{0+1+2}^{(\phi_1,\phi_2)} t_\mathrm{p}\right),$$

$$U_0(\tau) = \exp\left(-i\widetilde{H}_0\tau\right),$$

where $\tau$ is the free precession time and $t_\mathrm{p}$ is the width of the rectangular MW pulse. Thus, the DQ-1 Ramsey signal with the $|0\rangle$ state as the initial and readout state is calculated as:

$$S_{DQ1}^{(\phi_1,\phi_2)}(\tau) = \mathrm{Tr}\left[|0\rangle\langle 0|U_{DQ1}^{(\phi_1,\phi_2)}|0\rangle\langle 0|U_{DQ1}^{(\phi_1,\phi_2)\dagger}\right]$$

Since the 2[nd]-pulse phase effect can be separated as:

$$U_{0+1+2}^{(\phi_1,\phi_2)} = G_z(\phi_1,\phi_2)U_{0+1+2}^{(0,0)}G_z^\dagger(\phi_1,\phi_2),$$

where $G_z(\phi_1,\phi_2) \equiv \exp(-i[\phi_1 P_{+1} + \phi_2 P_{-1}])$ with the projection operators $P_m = |m\rangle\langle m|$, the signal can be simplified as:

$$\begin{aligned} S_{DQ}^{(\phi_1,\phi_2)}(\tau) &= \left|\mathrm{Tr}[P_0 U_{0+1+2} G^\dagger U_0 U_{0+1+2}]\right|^2 \\ &= \left|\mathrm{Tr}\left[G^\dagger(\Phi_1,\Phi_2)\, V(t_\mathrm{p})\right]\right|^2 \end{aligned}$$

with an effective operator representing rotation by the pulses and projection for readout

$$V(t_\mathrm{p}) \equiv U_{0+1+2}^{(0,0)}(t_\mathrm{p})|0\rangle\langle 0|U_{0+1+2}^{(0,0)}(t_\mathrm{p}),$$

and effective phase parameters $\Phi_j = \phi_j + \Delta_j\tau$ . In consequence, the Ramsey signal is decomposed into a series of interferometric terms expressed with the matrix elements $v_{i,j} = \langle i|V(t_\mathrm{p})|j\rangle$ in the $m_s$ basis:

$$\begin{aligned} S_{DQ\ (t_\mathrm{p})}^{(\phi_1,\phi_2)}(\tau) &= \left|v_{0,0} + e^{i\Phi_1}v_{1,1} + e^{i\Phi_2}v_{-1,-1}\right|^2 \\ &= \sum_m^{\{0,\pm1\}} \left|v_{m,m}\right|^2 + 2\Re\left[e^{i\Phi_1}v_{0,0}^* v_{1,1}\right] + 2\Re\left[e^{i\Phi_2}v_{0,0}^* v_{-1,-1}\right] \\ &\quad + 2\Re\left[e^{2i\Phi_\mathrm{d}}v_{1,1}v_{-1,-1}^*\right] \end{aligned}$$

Thus, the variables in Eq. (3) are explicitly described as:

$$c_0 = \sum_m^{\{0,\pm1\}} \left|v_{m,m}\right|^2$$

$$a_{2\mathrm{d}} = 2\left|v_{1,1}v_{-1,-1}^*\right|$$

$$2\psi_\mathrm{d} = \arg\left(v_{1,1}^* v_{-1,-1}\right)$$

$$a_m = 2\left|v_{0,0}v_{m,m}^*\right|\ (m = \pm1)$$

$$\psi_m = \arg\left(v_{0,0}v_{m,m}^*\right)\ (m = \pm1).$$

**APPENDIX B: Implementation of the DQ 4-Ramsey sequence in a mixing configuration**

Fig. 7 details the mixing configuration used to implement the DQ 4-Ramsey sequence shown in Fig. 2(e). As illustrated in Fig. 7(a), a SG provides a continuous microwave local oscillator (LO) signal at $\omega_\mathrm{LO}/2\pi \approx 2.87$ GHz, corresponding to the zero-field splitting. Simultaneously, an AWG synthesizes intermediate frequency (IF) signals at $\omega_\mathrm{IF}/2\pi \approx 70$ MHz, to address the Zeeman shift. A MW switch is employed to ensure high isolation during the MW-off periods.

To simultaneously drive all three $^{14}$N hyperfine-split states ($\omega_\mathrm{hf}/2\pi = 2.16$ MHz), the AWG generates multi-frequency signals for the in-phase (I) and quadrature (Q) ports of the IQ mixer. The IQ input signals $V_I(t)$ and $V_Q(t)$ are defined as:

$$V_I(t) = A_I(t)\sum_{m_I=-1}^{1}\cos\left((\omega_\mathrm{IF} - m_I\omega_\mathrm{hf})t + \phi_{m_I}\right)$$

$$V_Q(t) = A_Q(t)\sum_{m_I=-1}^{1}\sin\left((\omega_\mathrm{IF} - m_I\omega_\mathrm{hf})t + \phi_{m_I}\right)$$

where $A_I(t)$ and $A_Q(t)$ represent Gaussian amplitude profiles. The phase term $\phi_{m_I}$ is set to $\pi/2$ for the central component ($m_I = 0$) and 0 otherwise ($m_I = \pm1$). This phase shift is introduced to minimize peak power dissipation.

Upon mixing the LO and IF signals [Fig. 7(b)], the IQ mixer generates the radio-frequency (RF) output containing both the upper and lower sidebands ( $\omega_{\pm1}(m_I) = \omega_\mathrm{LO} \pm (\omega_\mathrm{IF} - m_I\omega_\mathrm{hf})$. For each hyperfine component $m_I$, the mixed output $V_\mathrm{RF}(t)$ can be described as:

$$\begin{aligned} V_\mathrm{RF}(t) &= A_I(t)\cos(\omega_\mathrm{LO}t)\cos\left((\omega_\mathrm{IF} - m_I\omega_\mathrm{hf})t\right) \\ &\quad + A_Q(t)\sin(\omega_\mathrm{LO}t)\sin((\omega_\mathrm{IF} - m_I\omega_\mathrm{hf})t) \\ &= \sum_{m=\pm1}\frac{A_I(t) - mA_Q(t)}{2}\cos(\omega_m(m_I)t). \end{aligned}$$

By controlling the amplitudes $A_I(t)$ and $A_Q(t)$ , this configuration allows for the precise generation of the $\omega_{\pm1}$ sidebands required for the DQ transitions. Appropriate modulation of the amplitude polarities that shifts the MW phase by $\pi$ cab implement the DQ 4-Ramsey sequence. Fig. 7(c) illustrates several options for transition selection depending on the chosen RF amplitudes. TABLE III shows the idealized amplitude combinations for the standard DQ 4-Ramsey sequence, where the labels and phase combinations correspond to those in TABLES I and II. These amplitudes can be flexibly adjusted to compensate for an RF power mismatch.

TABLE III. Normalized amplitude combinations on the second pulse for the standard DQ 4-Ramsey sequence. For all sequences, the amplitude of the first pulse is fixed at $A_I = 1$ and $A_Q = 0$.

| Label | A | B | C | D |
|---|---|---|---|---|
| $A_I$ | +1 | 0 | −1 | 0 |
| $A_Q$ | 0 | +1 | 0 | −1 |
| $A_I - A_Q$ for $\omega_{+1}$ | +1 | −1 | −1 | +1 |
| $A_I + A_Q$ for $\omega_{-1}$ | +1 | +1 | −1 | −1 |

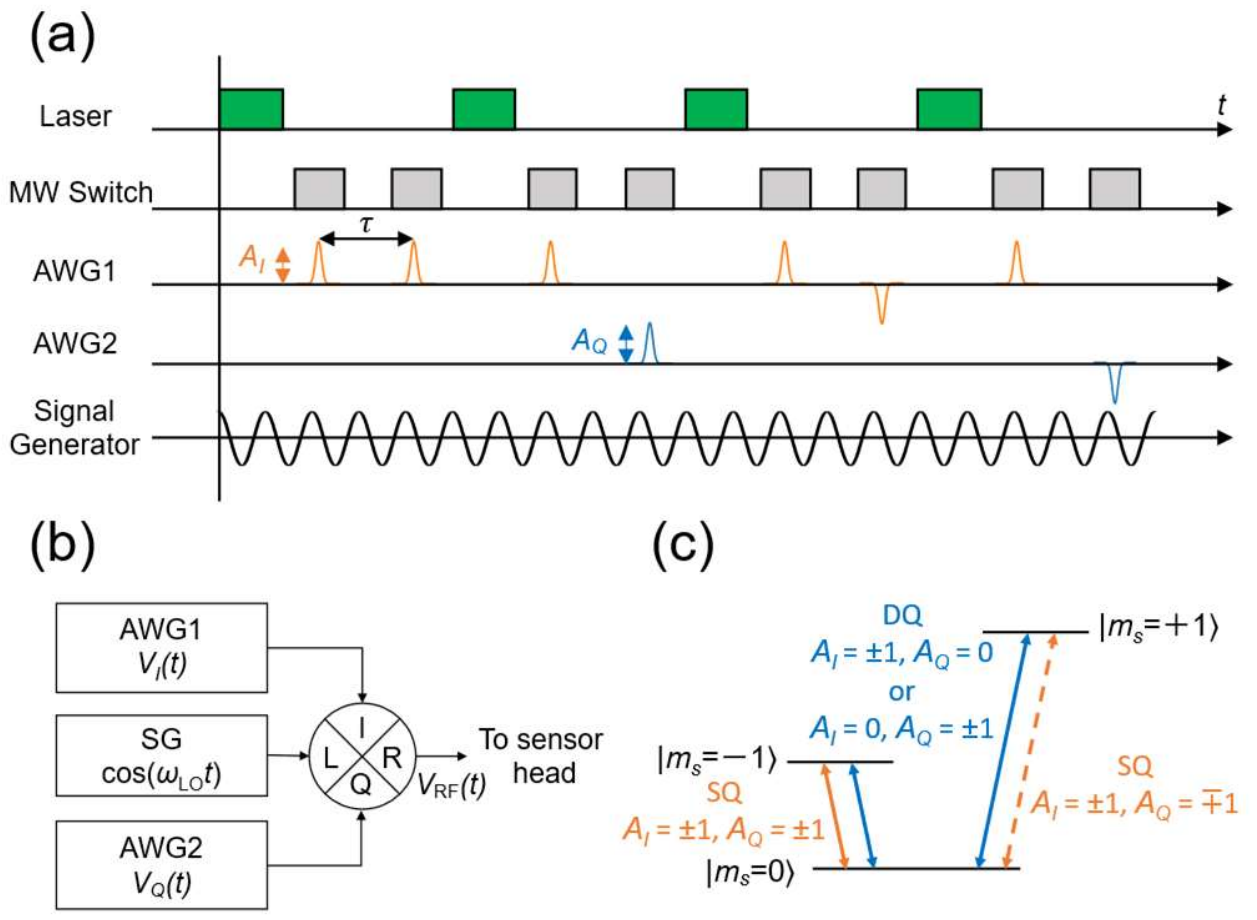

FIG. 7. Schematic of the MW and RF mixing configuration. (a) Timing diagram of the DQ 4-Ramsey sequence utilizing the mixing setup. (b) Block diagram of the MW and RF systems. (c) Drive frequency selection. Adjusting the amplitudes $A_I$ and $A_Q$ enables the selective driving of either SQ (orange) and DQ (blue) transitions.